\documentclass[11pt,aps,nofootinbib, notitlepage, onecolumn, a4paper,tightenlines,prl,longbibliography]{revtex4-1}
\usepackage[english]{babel}
\usepackage[utf8]{inputenc}
\usepackage{lmodern}
\usepackage[T1]{fontenc}
\usepackage{textcomp}
\usepackage{amsmath,amssymb,lmodern,amsthm}
\usepackage{mathtools}
\usepackage{graphicx}
\usepackage{mathrsfs}
\usepackage{hyperref}
\usepackage{url}
\usepackage{array}
\usepackage{tensor}
\usepackage{pifont}
\usepackage[toc,page]{appendix}
\usepackage{graphicx}
\usepackage[left=2.5cm,right=2.5cm,top=3cm,bottom=3cm]{geometry}
\usepackage{enumitem}
\usepackage{empheq}
\usepackage{cleveref}

\theoremstyle{plain}
\usepackage{xargs}
\usepackage{xcolor}
\hypersetup{
	colorlinks=true,
	urlcolor=blue,
	linkcolor={violet!50!black},
	citecolor={green!40!black}
}
\newtheorem{thm}{Theorem}[]
\theoremstyle{definition}

\newtheorem*{crit*}{Criterion}
\newtheorem{crit}{Criterion}[]
\newtheorem{example}{Example}[]

\renewcommand{\iff}{\Longleftrightarrow}
\newcommand{\liff}{\Longleftrightarrow}
\newcommand{\lied}{\pounds}
\newcommand{\defeq}{\ \vcentcolon=\ }
\newcommand{\defeqs}{\ \stackrel{\scri}{\vcentcolon=}\ }

\newcommand{\scri}{\mathscr{J}}

\newcommand{\pt}[2]{\tensor{\hat{#1}}{#2}}
\newcommand{\ctr}[3]{\tensor[{#1}]{#2}{#3}}
\newcommand{\ct}[2]{\tensor{{#1}}{#2}}

\newcommand{\cts}[2]{\tensor{\overline{#1}}{#2}}

\newcommand{\cs}[1]{#1}

\newcommand{\ps}[1]{\hat{#1}}

\newcommand{\df}[1]{\text{d}#1}

\newcommand{\prn}[1]{\left(#1\right)}

\newcommand{\brkt}[1]{\left[#1\right]}

\newcommand{\eqs}{\ \stackrel{\scri}{=}\ }
\newcommand{\eqsopen}{\ \stackrel{\Delta}{=}\ }

\newcommand{\cd}[1]{\tensor{\nabla}{#1}}

\newcommand{\commute}[2]{\left[#1,#2\right]}

\newcommand{\Q}{\mathcal{Q}}
\newcommand{\T}{\mathcal{T}}
\newcommand{\D}{\mathcal{D}}
\newcommand{\W}{\mathcal{W}}

\newcommand{\Pc}{\mathcal{P}}

\newcommand{\ms}[1]{\ct{h}{#1}}

\newcolumntype{M}[1]{>{\centering\arraybackslash}m{#1}}
\newcolumntype{N}{@{}m{0pt}@{}}

\def\be{\begin{equation}}
\def\ee{\end{equation}}
\def\bea{\begin{eqnarray}}
\def\eea{\end{eqnarray}}
\def\bean{\begin{eqnarray*}}
\def\eean{\end{eqnarray*}}

\newcounter{mnotecount}[section]

\renewcommand{\themnotecount}{\thesection.\arabic{mnotecount}}
\newcommand{\mnote}[1]
{\protect{\stepcounter{mnotecount}}$^{\mbox{\footnotesize
$
\bullet$\themnotecount}}$ \marginpar{
\raggedright\tiny\em
$\!\!\!\!\!\!\,\bullet$\themnotecount: #1} }

\makeatletter
\def\Dated@name{\normalsize} 
\let\@fnsymbol\@roman
\makeatother

\begin{document}

\title{\Large Gravitational radiation condition at infinity\\ with a positive cosmological constant}
	
\author{Francisco Fernández-Álvarez}
\email[Email address: ]{francisco.fernandez@ehu.eus} 
\author{José M. M. Senovilla} 
\email[Email address: ]{josemm.senovilla@ehu.eus}
\affiliation{Departamento de Física Teórica e Historia de la Ciencia,\\ Universidad del País Vasco UPV/EHU\\ Apartado 644, 48080 Bilbao, Spain }

\date{\today{}}

	\begin{abstract}
		Gravitational waves have been directly detected and astronomical observations indicate that our Universe has a positive cosmological constant $\Lambda$. Nevertheless, a theoretical gauge-invariant notion of gravitational waves arriving at infinity 
(escaping from the space-time) 
in the presence of a positive $\Lambda$ has been elusive. We find the answer to this long-standing gravitational puzzle and present a geometric, gauge-invariant radiation condition at infinity.
	\end{abstract}
\maketitle
	
	\section{Introduction}

	 Physics entered the new
era of gravitational-wave astronomy in 2016, following the announcement of the first gravitational-wave direct detection ever \cite{LigoVirgoPRL}. This was the successful culmination of a scientific adventure launched theoretically 
by Einstein a century before,
having the celebrated quadrupole formula \cite{Einstein18} in the weak-field limit as a solid first step, then followed by uncertainties
whether gravitational waves constitute a feature of the full non-linear theory, finally settled with various theoretical developments in the 1950-60's \cite{Trautman58,Pirani57,Bel1962,Bondi1962,Sachs1962,Penrose62}, see \cite{Zakharov}, and the discovery and analysis of the binary pulsar PSR B1913+16 \cite{Damour2015}.
To analyze isolated systems one focuses on regions of the space-time far away from the radiating sources, which formally correspond to `infinity'. The conformal geometric representation of these asymptotic regions by Penrose \cite{Penrose65} was particularly pioneering: carried out in full non-linear General Relativity in a covariant manner \cite{Geroch1977}, it contributed to dispel remaining doubts about the theoretical description of gravitational waves. 

While the conformal completion \cite{Penrose65} can be built for any value of the cosmological constant $ \Lambda $, its relationship with Bondi's fundamental quantities (news function and energy-momentum) \cite{Bondi1962} has only been established in the asymptotically flat case with $ \Lambda =0$ \cite{Geroch1977}. 
Observational data \cite{Riess1998,Perlmutter1999} reveal, however, that we inhabit an accelerated-expanding universe. This empirical fact evince the presence of a positive (bare or effective) cosmological constant. 
Thus, annoyingly we do not have a rigorous theoretical description of radiation escaping to infinity in the presence of a positive $\Lambda$, {\em no matter how tiny $\Lambda$ may be}. Signs of attention to this situation date back to \cite{Penrose2011}, and were amplified in \cite{Ashtekar2014} where the predicament was clearly presented. 
Some advances have been made \cite{Szabados2015,Ashtekar2015,Chrusciel2016,Saw2016,He2016,Compere2019,Poole2019,Virmani2019} (see \cite{Ashtekar2017,Szabados2019} for reviews), usually trying to adapt techniques from the $ \Lambda=0 $ case to the new scenario. Frustratingly, we still lack a fully satisfactory solution. One of the challenging difficulties is to understand and describe unambiguously the directional dependence that emerges when one approaches infinity in different lightlike directions \cite{Krtous2004}. Not to mention the absence of an asymptotic universal structure of infinity ---which does exist for $ \Lambda=0 $, allowing to isolate the two degrees of freedom associated to gravitational radiation \cite{Ashtekar81}.

In summary, the next question remains open: \emph{How to tell when a space-time with positive cosmological constant contains gravitational radiation arriving at infinity?} In this letter, we answer this fundamental question which underlies any other hypothetical deeper characterisation, such as a formula for the energy carried away by the waves from an isolated source or the definition of a mass-energy.  We do it by taking a fully new perspective of the problem, different from the methods used so far in the literature. Namely, we ground our investigation in studying tidal effects, motivated by the nature of the gravitational field and of actual gravitational-wave measurements. Our approach is supported by its successful application to the well-established asymptotically flat case, in which we recently put it at test \cite{Fernandez-Alvarez_Senovilla20}: we demonstrated that our tidal approach is fully equivalent, in a precise sense, to the traditional scheme with $\Lambda =0$. By these tidal means, hitherto we arrive at a satisfactory \emph{radiation condition at infinity} in the presence of a positive cosmological constant. As far as we know, this is the first such criterion. 

	\section{Conformal space-time and superenergy}
		The suitable structure for the study of asymptotic gravitational radiation is the boundary $ \scri $ of a conformal completion\footnote{The metric signature is $ (-,+,+,+) $, and the curvature tensor is defined by $ (\cd{_\alpha}\cd{_\beta}-\cd{_\beta}\cd{_\alpha})\ct{v}{_\gamma} = \ct{R}{_{\alpha\beta\gamma}^\mu} \ct{v}{_\mu}$, where $ \cd{_\alpha} $ is the covariant derivative on $ \left(\cs{M},\ct{g}{_{\alpha\beta}} \right) $.
} $ \prn{\cs{M},\ct{g}{_{\alpha\beta}}} $ of any physical space-time $ \prn{\ps{M},\pt{g}{_{\alpha\beta}}} $ \cite{Penrose65}, where $\ct{g}{_{\alpha\beta}}= \Omega^2 \pt{g}{_{\alpha\beta}} $ on $\hat M$, the conformal factor $\Omega$ is strictly positive on $ \ps{M} $ and `infinity' lies at $ \scri :=\{\Omega=0 \}$; for further details see e.g. \cite{Geroch1977,Kroon}. For $ \Lambda>0 $, $\scri$ is a three-dimensional manifold endowed with a Riemannian 
metric\footnote{Latin indices $ a=1,2,3 $ are intrinsic to $ \scri $.} $ \ms{_{ab}}$ inherited from $ \prn{\cs{M},\ct{g}{_{\alpha\beta}}} $. Generically $\scri$ is not connected, its components can be divided into `future' and `past' denoted by $\scri^\pm$ respectively; our discussion is valid for any of them, but we will sometimes concentrate on $\scri^+$ for the sake of concreteness (escaping radiation). 
The normal $ \ct{N}{_\alpha}\defeq \cd{_\alpha}\Omega $ to the $ \Omega=\text{constant} $ hypersurfaces  is future timelike on a neighbourhood of $\scri$, for $ -N^2\defeq\ct{N}{_\alpha}\ct{N}{^\alpha}\eqs -\Lambda/3  $. In such a neighbourhood one can normalise it $ \ct{n}{_\alpha}\defeq N^{-1}\ct{N}{_\alpha} $. 
There is a gauge freedom consisting on conformal rescaling of $ \ct{g}{_{\alpha\beta}} $, $\Omega \rightarrow \Omega \omega$ with $ \omega >0 $. 
It is common to restrict $ \omega $ such that $ \cd{_\alpha}\ct{N}{_\beta}\eqs 0$, a `divergence-free' gauge that we adopt. Still, a huge gauge freedom persists, as any $ \omega $ satisfying $ \lied_{\vec{N}}\omega\eqs 0 $ is still allowed. 
Our main results are fully gauge independent.

		In order to motivate our novel approach, let us throw a quick sight into the electromagnetic (EM) field, whose energy-momentum tensor
		 we denote by $ \ctr{^{\scriptscriptstyle EM}}{T}{_{\mu\nu}} $. For any unit, future-pointing, vector $ \ct{v}{^\alpha} $, the four-momentum vector relative to $v^\alpha$  is defined as $ \ctr{^{\scriptscriptstyle EM}}{P}{^\alpha}\defeq-\ctr{^{\scriptscriptstyle EM}}{T}{^\alpha_\mu} \ct{v}{^\mu}$. This is always causal and future, its $ \ct{v}{^\alpha} $-component gives the EM energy density measured by $v^\alpha$, while the remaining, three-dimensional, spatial vector is the Poynting vector. The latter points in the spatial direction along which EM energy propagates according to the observer described by $ \ct{v}{^\alpha} $. {\em Null EM fields} are characterized by having a non-zero Poynting vector for all possible $v^\alpha$.
		
		In General Relativity, things are more complicated. A great obstacle is the absence of a local notion of energy-momentum associated to the gravitational field, although there are quasilocal definitions \cite{Szabados2004}. Instead of working at the energy-density level 
one can work with tidal energies. At this level the \emph{Bel-Robinson tensor} \cite{Bel1958} is defined
			\begin{equation}\label{BR-tensor}
			\ct{\T}{_{\alpha\beta\gamma\delta}}\defeq \ct{C}{_{\alpha\mu\gamma}^\nu}\ct{C}{_{\delta\nu\beta}^\mu} + \ctr{^*}{C}{_{\alpha\mu\gamma}^\nu}\ctr{^*}{C}{_{\delta\nu\beta}^\mu}
			\end{equation}	
where $\ct{C}{_{\alpha\mu\gamma}^\nu}$ is the Weyl tensor and $\ctr{^*}{C}{_{\alpha\mu\gamma}^\nu}$ its Hodge dual.
		$\ct{\T}{_{\alpha\beta\gamma\delta}}$ is a totally symmetric, traceless, conformally invariant tensor whose physical units are $ ML^{-3}T^{-2} $ \cite{Teyssandier1999,Senovilla2000,Szabados2004} and carries information related to the tidal nature of the gravitational field. The objects defined with this tensor are usually referred to as \emph{superenergy} quantities. 
		Given an observer $ \ct{v}{^\alpha} $ as before, one defines the associated supermomentum
			\begin{equation}\label{supermomentum}
			\ct{P}{^\alpha} \defeq -  \ct{v}{^\beta}\ct{v}{^\gamma}\ct{v}{^{\delta}}\ct{\T}{^\alpha_{\beta\gamma\delta}}= \cs{W}\ct{v}{^\alpha} + \ct{S}{^\alpha}	
			\end{equation}
		where $ \cs{W} $ and $ \ct{S}{^\alpha} $ ($S^\alpha v_\alpha =0$) are the superenergy density and super-Poynting vector relative to $v^\alpha$
			\begin{align}
			\cs{W}&\defeq \ct{v}{^\alpha}\ct{v}{^\beta}\ct{v}{^\gamma}\ct{v}{^\delta}\ct{\T}{_{\alpha\beta\gamma\delta}} \geq 0 \quad ,\label{W}\\
			\ct{S}{^\alpha} &\defeq -(\delta_\mu^\alpha + \ct{v}{^\alpha}\ct{v}{_\mu})\ct{v}{^\beta}\ct{v}{^\gamma}\ct{v}{^{\delta}}\ct{\T}{^\mu_{\beta\gamma\delta}} \label{P}\quad.	
			\end{align}
		The Bel-Robinson tensor has many relevant properties \cite{Bel1962,BonillaSenovilla97,Senovilla2000} analogous to the ones fulfilled by $ \ctr{^{\scriptscriptstyle EM}}{T}{_{\mu\nu}} $.
Among them, a dominant property \cite{Senovilla2000} ensuring that $ \ct{P}{^\alpha} $ is causal and future pointing. In analogy with the null EM fields, Bel \cite{Bel1962} proposed to define a state of intrinsic gravitational radiation at a point $ q $ when $ \ct{S}{^\alpha}|_q \neq 0$ for all possible $\ct{v}{^\alpha}  $. This is a local, observer-independent statement. 

		\section{The radiation criterion at $\scri$}
		We aim at an observer-independent and gauge-invariant description of gravitational radiation at the conformal boundary $\scri$. For such purpose, and recalling that we are interested in studying tidal effects, first we have to find a good definition of \emph{asymptotic supermomentum}. Something as (\ref{supermomentum}) does not work, as
we know that the Weyl tensor vanishes at $ \scri $ \cite{Geroch1977,Kroon}. Nevertheless, the rescaled Weyl tensor $ \ct{d}{_{\alpha\beta\gamma}^\delta}\defeq \Omega^{-1}\ct{C}{_{\alpha\beta\gamma}^\delta} $ is regular at infinity. Consequently, to carry out an asymptotic study in conformal space-time, it is natural to define a \emph{rescaled Bel-Robinson tensor},
			\begin{equation}
			\ct{\D}{_{\alpha\beta\gamma\delta}} \defeq \Omega^{-2}\ct{\T}{_{\alpha\beta\gamma\delta}} =
			  \ct{d}{_{\alpha\mu\gamma}^\nu}\ct{d}{_{\delta\nu\beta}^\mu} + \ctr{^*}{d}{_{\alpha\mu\gamma}^\nu}\ctr{^*}{d}{_{\delta\nu\beta}^\mu} 
			\end{equation}
which shares most of the properties of $ \ct{\T}{_{\alpha\beta\gamma\delta}} $ and is regular and --in general-- different from zero at $ \scri $.
Notice that the normal $ \ct{N}{_\alpha}|_\scri $ 
defines a privileged `asymptotic observer' that is selected by the geometry itself. Hence, it is natural to define the \emph{asymptotic supermomentum}  
			\begin{equation}\label{super-momentum}
			\ct{p}{^\alpha}\defeq -\ct{N}{^\mu}\ct{N}{^\nu}\ct{N}{^\rho}\ct{\D}{^\alpha_{\mu\nu\rho}}
			\end{equation}
and its \emph{canonical} version
\begin{equation}\label{canonical-momentum}
			\ct{\Pc}{^\alpha}\defeq -\ct{n}{^\mu}\ct{n}{^\nu}\ct{n}{^\rho}\ct{\D}{^\alpha_{\mu\nu\rho}}\quad.
			\end{equation}
Obviously, these two vector fields are proportional in a neighbourhood of $\scri$, $\ct{\Pc}{^\alpha}=N^{-3} p^\alpha$, and all the properties listed below hold for both of them. However, (\ref{canonical-momentum}) is unsuitable for comparison with the $\Lambda=0$ case, for which we will use (\ref{super-momentum}) later on.
Some important properties of these vector fields are:
			\begin{enumerate}
				\item \label{P-future-property} $  \ct{\Pc}{^\alpha}$ is causal and future pointing at and around $\scri$.
				\item Under gauge transformations, it changes at $\scri$ as 
			\begin{equation}\label{gauge-P}
			\ct{\Pc}{^\alpha} \stackrel{\scri}{\rightarrow}\omega^{-7}\ct{\Pc}{^\alpha} \quad.
			\end{equation}
			\item \label{divergence-free-property} If the energy-momentum tensor of the physical space-time $(\hat M,\hat g_{\mu\nu})$ behaves near $\scri$ as $ \pt{T}{_{\alpha\beta}}|_\scri \sim \mathcal{O}(\Omega^3)$ (which includes the vacuum case $ \pt{T}{_{\alpha\beta}}=0 $), then 
			$$ 
			\cd{_\mu}\ct{\Pc}{^\mu}\eqs 0 \quad. 
			$$
			\end{enumerate}
			
		Let $ \{\ct{n}{^\alpha}|_\scri ,\ct{e}{^\alpha_a}\} $ 
be an orthonormal basis 
of $ \cs{M} $ at $ \scri $. 
		The canonical asymptotic supermomentum decomposes as
			\begin{equation}
				\ct{\Pc}{^\alpha}\eqs \cs{\W}\ct{n}{^\alpha} + \cts{\Pc}{^\alpha}\eqs\cs{\W}\ct{n}{^\alpha} + \cts{\Pc}{^a}\ct{e}{^\alpha_a} \quad,
			\end{equation}
		where,
			\begin{align}
			\cs{\W} &\defeqs -\ct{n}{_\mu}\ct{\Pc}{^\mu}\geq 0 \quad \label{superenergyDef},\\
			\cts{\Pc}{^\alpha} &\defeqs -(\delta^\alpha_\mu +n^\alpha n_\mu) \cs{\Pc}^\mu\quad,  \quad \cts{\Pc}{^\alpha} n_\alpha =0 \quad, \label{superPoyntingDef}\quad
			\end{align}
		are the asymptotic superenergy and super-Poynting vector field, respectively. At this stage, we can already introduce the gravitational-radiation condition.
		
		\begin{crit}[Asymptotic gravitational-radiation condition with $ \Lambda>0 $]\label{criterionGlobal}
			Consider a three-dimensional open connected subset $ \Delta \subset \scri $.
			There is no radiation on $ \Delta $  if and only if the asymptotic super-Poynting vanishes there 
			\begin{equation*}
			 \cts{\Pc}{^\alpha}\eqsopen 0 \iff \text{No gravitational radiation on }\Delta .
			\end{equation*}
		\end{crit}		
Observe that an equivalent statement is that the super-momentum points along the normal $N^\alpha$ at $\scri$. This allows to give two alternative but equivalent formulations of the criterion which will be valuable for later comparison. 
\begin{itemize}
\item  No gravitational radiation on $\Delta\subset \scri \iff \ct{p}{^\alpha}$ is orthogonal to all surfaces 
within $\Delta$.
\item No gravitational radiation on $\Delta\subset \scri \iff N^\alpha|_\scri$ is a principal vector (in the sense of Pirani, i. e., those lying in the intersection of two principal planes, see \cite{Pirani57,Bel1962,Ferrando1997}) of $\ct{d}{_{\alpha\beta\gamma}^\delta}|_\Delta$ .
\end{itemize}
		
%
{\bf Remarks:}	
		\begin{enumerate}
		\item
			This characterisation is gauge invariant, as follows from (\ref{gauge-P}).
		\item
			In terms of the electric, $ \ct{D}{_{ab}}\defeqs \ct{e}{^\alpha_a}\ct{e}{^\beta_b}\ct{n}{^\mu}\ct{n}{^\nu}\ct{d}{_{\mu\alpha\nu\beta}} $, and magnetic, $ \ct{C}{_{ab}}\defeqs \ct{e}{^\alpha_a}\ct{e}{^\beta_b}\ct{n}{^\mu}\ct{n}{^\nu}\ctr{^*}{d}{_{\mu\alpha\nu\beta}} $, parts of the rescaled Weyl tensor, the asymptotic superenergy and super-Poynting take the form \cite{Bel1962,Maartens1998}
				\begin{align}
			\W&\eqs  \ct{D}{_{ab}} \ct{D}{^{ab}}+  \ct{C}{_{ab}} \ct{C}{^{ab}}\quad,\\
			 \label{poyntingCommutator}
			 \cts{\Pc}{^a}&\eqs \commute{C}{D}_{rs}\ct{\epsilon}{^{rsa}}\eqs 2\ct{C}{_r^t}\ct{D}{_ {ts}}\ct{\epsilon}{^{rsa}}\quad,
			\end{align}
			where $ \ct{\epsilon}{_{abc}} $ is the volume 3-form of $(\scri ,h_{ab})$. This means that there are no gravitational waves at $ \scri $ if and only if $ \ct{C}{_{ab}} $ and $ \ct{D}{_{ab}} $ commute. A necessary condition for this to occur is that $\ct{d}{_{\alpha\beta\gamma}^\delta}|_\scri$ be of Petrov type I or D \cite{Bel1962,Alfonso2008}. Observe, though, that this is not sufficient and these two types can contain asymptotic radiation -- later on, we will present one example (\Cref{example-Cmetric}). 
		\item
			The existence of radiation depends on the interplay between $ \ct{D}{_{ab}} $ and $ \ct{C}{_{ab}} $, and cannot be determined with only one of them --letting aside the trivial case in which either vanishes. The Cotton-York tensor of $(\scri,h_{ab})$ can be shown to be $-\sqrt{\Lambda/3}\, C_{ab}$, and thus $C_{ab}$ is determined by the intrinsic geometry of $\scri$. Hence, this intrinsic geometry is not enough to encode the presence of asymptotic radiation, and one needs to bring $D_{ab}$ into the picture. This agrees with the next remark.
		\item \label{remark4}
		A fundamental result \cite{Friedrich1986a,Friedrich1986b} states that a solution of the $\Lambda$-vacuum Einstein field equations is fully determined by initial/final data consisting of the conformal class of a 3-dimensional Riemmanian manifold $(\scri,h_{ab})$ plus a traceless and divergence-free tensor $D_{ab}$. Hence, the existence or not of radiation \underline{is encoded} in $(\scri,h_{ab},D_{ab})$, and our \Cref{criterionGlobal} entails this fact neatly and for the first time.
		
		\item 
		The condition as stated is computation friendly, and can be easily implemented in algebraic computing programs.
	\end{enumerate}
	The radiation detected with the above criterion at $\scri^+$ is always `escaping' from the physical space-time, and $\cts{\Pc}{^\alpha}$ points in its spatial direction of propagation there. However, at this stage, one cannot say if the radiation is originated by 
	 an isolated source, as there can be other pieces of radiation coming to $\scri^+$ from beyond the cosmological horizon of such isolated source \cite{Ashtekar2014,Ashtekar2019}. More on this later.
		
	In order to show that \Cref{criterionGlobal} is reliable, we first compare with the $ \Lambda=0 $ case and then present relevant examples supporting it ---with and without radiation.
	\section{Comparison with the asymptotically flat case ($\Lambda =0$)}
		The first distinguishable property of the asymptotically flat scenario is that $\scri$ is a lightlike hypersurface, as the normal $\ct{N}{_\alpha} $ is lightlike and thus $\ct{N}{^\alpha} $ is tangent to $\scri$---and cannot be normalized. Also, the topology of $\scri$ is determined to be $ \mathbb{R}\times\mathbb{S}^2 $ \cite{Penrose65,Geroch1977}.
		In this $ \Lambda=0 $ case the presence of \emph{escaping} gravitational radiation --i.e. transversal to $ \scri^+ $-- is successfully determined by the \emph{news tensor} \cite{Geroch1977,Ashtekar81} $ \ct{N}{_{ab}} $---equivalent to the Bondi news function \cite{Bondi1962}  in a particular gauge---: a symmetric, traceless, gauge-invariant tensor field on $ \scri^+$ orthogonal to its null generators. Specifically, the standard criterion for the existence of radiation on an open portion $ \Delta \subset \scri^+$ with same topology as $ \scri^+ $ reads for $\Lambda =0$ \cite{Geroch1977,Ashtekar81}
			\begin{equation*}
				\ct{N}{_{ab}}\eqsopen 0 \iff \text{No gravitational radiation on }\Delta .
			\end{equation*}
		Actually, using the results in \cite{Fernandez-Alvarez_Senovilla20} we can show that this criterion agrees with ours. 
		To prove it we must use the analog of $\ct{p}{^\alpha}$ when $\Lambda =0$, which was introduced in \cite{Fernandez-Alvarez_Senovilla20} as 
			\begin{equation}\label{radiant-supermomentum}
			\ct{\Q}{^\alpha}\defeq -\ct{N}{^\mu}\ct{N}{^\nu}\ct{N}{^\rho}\ct{\D}{^\alpha_{\mu\nu\rho}}\quad (\Lambda=0).
			\end{equation}
	Observe that this is formally the same definition as (\ref{super-momentum}); in fact, for space-times in which the limit $\Lambda\rightarrow 0$ makes sense\footnote{An example where such a limit does not exist is the $a\rightarrow \infty$ limit of Kerr-de Sitter \cite{MPS,MPS1}.} one has 
			\begin{equation}\label{eq:limit}
				\lim_{\Lambda \rightarrow 0} \ct{p}{^\alpha}\vert_{\scri} = \ct{\Q}{^\alpha}|_{\scri}\quad.
			\end{equation}
	 $\ct{\Q}{^\alpha}|_{\scri}$ is
		observer independent, geometrically distinguished and, due to known properties \cite{Senovilla2000,Bergqvist2004},
		is future {\em ligthlike}.
		This is why we call it the 
		\emph{asymptotic radiant supermomentum}.
		In \cite{Fernandez-Alvarez_Senovilla20} we proved 
			\begin{equation*}
			\ct{N}{_{ab}}\eqsopen  0 \quad\liff\quad \Q{^\alpha} \eqsopen  0\quad.
			\end{equation*}
		In simpler words, the standard radiation condition for $\Lambda=0$ can be re-stated as
		\begin{equation*}
			  \Q{^\alpha} \eqsopen 0 \iff \text{No gravitational radiation on }\Delta.
			\end{equation*}
		It might seem at first sight that there is a slight difference between this and  \Cref{criterionGlobal}, but this is only apparent: $\Q{^\alpha}|_{\scri}$ admits decompositions into a vector field $\cts{\Q}{^\alpha}$ tangent to $\scri$ --`radiant super-Poynting'-- and a transverse component. The vanishing of the former on $ \Delta $ is equivalent to the vanishing of the whole $\Q{^\alpha}$ on that open portion \cite{Fernandez-Alvarez_Senovilla20}.
		
		To reinforce the agreement between  \Cref{criterionGlobal} for $\Lambda >0$ and the classical criterion for $\Lambda=0$ we rewrite the latter in the following two equivalent forms:
	\begin{itemize}
\item  No gravitational radiation on $\Delta\subset \scri \iff \Q^\alpha$ is orthogonal to all surfaces within $\Delta$.
\item No gravitational radiation on $\Delta\subset \scri \iff N^\alpha|_\scri$ is a principal vector of $\ct{d}{_{\alpha\beta\gamma}^\delta}|_\Delta$.
\end{itemize}
	These are the same characterizations as those given previously for the case $\Lambda >0$.	
		 
		No equivalent of the news tensor has been found when $ \Lambda>0 $ yet, which enhances the relevance and versatility of \Cref{criterionGlobal}: the radiation \emph{escaping} from the physical space-time is determined by the asymptotic supermomentum --(\ref{super-momentum}) or (\ref{radiant-supermomentum})-- defined with the normal to $ \scri^+ $, no matter if $ \Lambda=0 $ or $ \Lambda>0 $. 
		
	\section{On `incoming radiation' at infinity}
		An important clarification is necessary: the lightlike character of $\scri^+$ when $\Lambda=0$ excludes any possible incoming waves, which would propagate tangentially to $ \scri^+ $, and thus the radiation detected by $N_{ab}$, or equivalently by $\Q^\alpha|_{\scri^+}$, is entirely due to an isolated system, in contrast with the $\Lambda >0$ case. This makes it clear that, in order to transfer all other important results from the $\Lambda=0$ case --such as energy loss, balance laws, etc-- to $\Lambda> 0$ a full control of `incoming radiation' is needed. There exists one proposal \cite{Ashtekar2019} that requires knowledge of the physical space-time,  but, in tune with Remark \ref{remark4} above, absence of incoming radiation must be encoded in $(\scri^+, h_{ab}, D_{ab})$, hence we aim at handling this problem using information solely from those data: it must involve $D_{ab}$ and the intrinsic geometry of $(\scri^+,h_{ab})$ ($C_{ab}$ for instance).
	
There are multiple conditions of that type, but to motivate our approach we present here one that is salient and useful. Gaining inspiration from the $\Lambda=0$ case, we can import formula (\ref{radiant-supermomentum}) to the $\Lambda>0$ situation and define asymptotic {\em radiant} super-momenta associated to every null vector field $k^\mu$ ($k^\mu n_\mu =-1$) on $\scri^+$:
\be\label{radiant-supermomenta}
{}^{(k)}\Q^\alpha \stackrel{{\scri^+}}{\defeq} -\ct{k}{^\mu}\ct{k}{^\nu}\ct{k}{^\rho}\ct{\D}{^\alpha_{\mu\nu\rho}} .
\ee
All these vector fields are future lightlike \cite{Senovilla2000,Bergqvist2004}, its vanishing for a particular $k^\mu$	is related to the absence of radiation in directions {\em transversal} to $k^\mu$. There is a unique unit vector field tangent to $\scri^+$ in the 2-planes defined by $k^\mu$ and $n^\mu$, given by $ \ct{r}{^\alpha}\stackrel{{\scri^+}}{\defeq} \ct{n}{^\alpha}-{k}{^\alpha}= r^ae^\alpha{}_a$.
Thus, an obvious thing to do is to require that ${}^{(k)}\Q^\alpha=0$ for some $k^\mu$ such that the corresponding $r^\mu$ points towards the region where the isolated source meets $\scri^+$. (This implies that $ {k}{^\alpha}  $ is a multiple principal null direction of $ \ct{d}{_{\alpha\beta\gamma}^\delta}|_{\scri^+} $).	
Condition ${}^{(k)}\Q^\alpha=0$ is of the type we are seeking, because one can prove \cite{Fernandez-Alvarez_Senovilla} its equivalence to the following relationship between $ \ct{D}{_{ab}} $ and $ \ct{C}{_{ab}} $:
\be\label{NIRC}
D_{ab}-\frac{1}{2}D_{ef}r^e r^f \left(3r_a r_b - h_{ab} \right)= r^d\epsilon_{ed(a} \left(C_{b)}{}^e +r_{b)} r^f C_f{}^e \right)
\ee
so that $ \ct{D}{_{ab}} $ is determined by $ \ct{C}{_{ab}} $ except for the one component $ \ct{D}{_{ef}}\ct{r}{^e}\ct{r}{^f} $.


Therefore, assuming the existence of a $k^\mu$ such that its radiant super-momenta (\ref{radiant-supermomenta}) vanishes looks like a good condition removing incoming radiation at $\scri^+$. This is further supported by the following interesting result that can be proven using (\ref{NIRC}) \cite{Fernandez-Alvarez_Senovilla}:
		\begin{thm}
			Assume that ${}^{(k)}\Q^\alpha=0$ for some given $k^\mu$  on $ \Delta \subset \scri^+ $. Then, $\cts{\Pc}{^\alpha}r_\alpha \leq 0$. Moreover,
				\begin{equation*}
					{}^{(\ell)}\Q^\alpha \eqsopen 0  \iff \cts{\Pc}{^\alpha}r_\alpha \eqsopen 0 \iff \cts{\Pc}{^\alpha}\eqsopen 0 \iff \text{No gravitational radiation on }\Delta
				\end{equation*}
where $\ell^\mu \defeqs n^\mu +r^\mu=2n^\mu -k^\mu$ is the unique lightlike vector field in the plane $\left<k,n\right>$ ($\ell^\mu n_\mu =-1$) other than $ \ct{k}{^\mu} $.
		\end{thm}

Observe that the vanishing of (\ref{radiant-supermomenta}) for an appropriate $k^\mu$ endows $(\scri^+,h_{ab},D_{ab})$ with further structure:  the congruence of curves tangent to $ \ct{r}{^\alpha} $  (compare \cite{Compere2020}).
Upon certain conditions this congruence of curves is shearless (and then, choosing the gauge, non-expanding too).  This permits to define a universal structure on $\scri^+$ together with a set of adapted asymptotic symmetries \cite{Fernandez-Alvarez_Senovilla}.
		

	\section{examples}
		\begin{example}
				As an obvious example, every space-time whose metric $h_{ab}$ on $ \scri $ is conformally flat has $ \ct{C}{_{ab}}\eqs 0 $ and, consequently, it does not contain gravitational radiation at infinity. Similarly, every space-time with $D_{ab}=0$ does not have radiation either. Likewise, conformally flat physical space-times (that is, with a vanishing Weyl tensor) have no radiation as $ \ct{d}{_{\alpha\beta\gamma}^\delta}\eqs 0$. 
Friedman-Lema\^{\i}tre-Robertson-Walker models admitting $\scri$ are relevant examples including, of course, de Sitter (dS) space-time.

		\end{example}
		\begin{example}
				If the physical space-time is spherically symmetric and the conformal completion respects the symmetry then $ \ct{C}{_{ab}} $ and $ \ct{D}{_{ab}} $ inherit the symmetry too, implying that in a preferred basis they are both diagonal, ergo they conmute, and therefore there is no gravitational radiation. This holds true also for all space-times with a 3-dimensional group of symmetries ---not only SO(3)--- with spacelike 2-dimensional orbits, as long as $\Omega$ is invariant by the group. Relevant examples are the generalized Kottler space-times \cite{Stephani2003,MPS} or the Vaidya metric.
		\end{example}

		\begin{example}
			The family of Kerr-de Sitter-like space-times comprises all vacuum space-times with positive cosmological constant admitting a conformal completion and whose Mars-Simon tensor (see e.g. \cite{Parrado2013,Mars2016,MPS} and references therein) with respect to a given Killing vector vanishes. These space-times have been studied in \cite{Mars2016}, and they include in particular Kerr-dS, Kerr-NUT-dS, or Taub-NUT-dS. The Killing vector induces a conformal Killing vector $Y^a$ on $\scri$ without fixed points and the explicit expressions for $ \ct{C}{_{ab}} $ and $ \ct{D}{_{ab}} $ were found to be both proportional to \cite{Mars2016}
				$$
				Y_a Y_b -\frac{1}{3} Y_c Y^c h_{ab} \quad.
				$$
			Hence, $\cts{\Pc}{^a} =0$ and these space-times contain no gravitational radiation at infinity according to our criterion, as expected.
		\end{example}

		\begin{example}\label{example-Cmetric}
			We deal now with the C-metric with $\Lambda\geq 0$, a solution which under certain range of parameters represents two accelerating black holes in a dS ($\Lambda >0$) or flat ($\Lambda =0$) background. Gravitational waves at infinity are thus expected; actually, the existence of a non-vanishing news tensor when $ \Lambda=0 $ was demonstrated in \cite{Ashtekar-Dray81}. For further insights on the properties of these metrics, see \cite{Griffiths-Podolsky2009}. 
			
			We use a recent version of this metric
			\cite{Podolsky2009}, in which the conformal space-time metric in a divergence-free gauge reads
			\begin{equation}
			\df{s^2}=\frac{1}{S}\prn{T\df{\tau^2}-\frac{1}{T}\df{q^2}+\frac{1}{S}\df{p^2}+S\df{\sigma^2}}\quad,
			\end{equation}
		where 
			\begin{align}
			T(q) &\defeq (\alpha^2-q^2)(1+2 mq) +\Lambda/3 
			\quad,\\
			S(p) &\defeq (1-p^2)(1-2\alpha mp) 
			\end{align}
			and the conformal factor is 
			$ \Omega^2\defeq \prn{q + \alpha p}^2/S $ ergo $\scri$ is defined by $q+\alpha p=0$. There are two Killing vectors: $\partial_\tau$ with $\mathbb{R}$-orbits, and $\partial_\sigma$ with cyclic orbits but conical singularities at $p=\pm 1$, so that $p\in (-1,1)$ and thus $S>0$ ---by assuming $2\alpha m <1$.  To describe a region containing $\scri$ we must choose $q \in (-\alpha,\alpha)$, 
			so that $T(q) >0$ in the chosen range, and the space-time is non-stationary there. The roots where $T(q)=0$ are horizons through which the metric can be regularly extended. Notice that these horizons do not intersect $\scri$.
		There are four constant parameters: $ \alpha $ (acceleration), $ m $ (mass), $ \Lambda $ and $ C $, the latter sometimes is used to remove one of the conical singularities and defines the range of the coordinate $ \sigma\in \left[0,2\pi C\right) $ \cite{Griffiths-Podolsky2009}. 
		
		The metric on $ \scri $ reads
			\begin{equation}
			h= \frac{1}{S}\prn{ (\alpha^2S+\Lambda/3) \df{\tau^2}+\frac{\Lambda}{3S(\alpha^2S+\Lambda/3)}\df{p^2}+S\df{\sigma^2}}\quad.
			\end{equation}
		This is clearly positive definite having a smooth limit when $\Lambda \rightarrow 0$ that leads to a degenerate $h_{ab}$.
		The electric and magnetic parts of the rescaled Weyl tensor on $ \scri $, in the coordinate basis, have as non-zero components for $\Lambda > 0$
			\begin{align}
				\ct{C}{_{\tau\sigma} }= \frac{3}{\Lambda}\alpha m S^{3/2} (3S\alpha^2+\Lambda), 
				\quad \ct{D}{_{\sigma\sigma}}=\frac{m}{\Lambda}S^{3/2} \prn{\Lambda+9S\alpha^2}, \\
				\ct{D}{_{\tau\tau}} = -\frac{m}{\Lambda}\sqrt{S} \prn{9S^2\alpha^4+5\Lambda S\alpha^2+\frac{2}{3}\Lambda^2},
				\quad \ct{D}{_{pp}} = \frac{m\Lambda}{\sqrt{S}\prn{\Lambda+3S\alpha^2}}
			\end{align}
hence, the asymptotic canonical super-Poynting vector and superenergy, represented in \cref{figure}, have the following expressions:
			\begin{equation}\label{super-PinC}
				\cts{\Pc}{^a}\eqs \sqrt{\frac{3}{\Lambda}}18\alpha m^2 S^{9/2} \prn{1+\frac{6}{\Lambda}S\alpha^2}\delta_{p}^a\quad,
			\end{equation}
			\begin{equation}\label{WinC}
				\cs{\W}\eqs 6m^2 S^3 \prn{1+\frac{54}{\Lambda^2}S^2\alpha^4+\frac{18}{\Lambda}S\alpha^2}\quad.
			\end{equation}
		 The super-Poynting vector field is different from zero everywhere on $ \scri $, thus meaning that gravitational radiation arrives there, as expected. 
		 \begin{figure}
		 	\centering
		 	\includegraphics[scale=0.4]{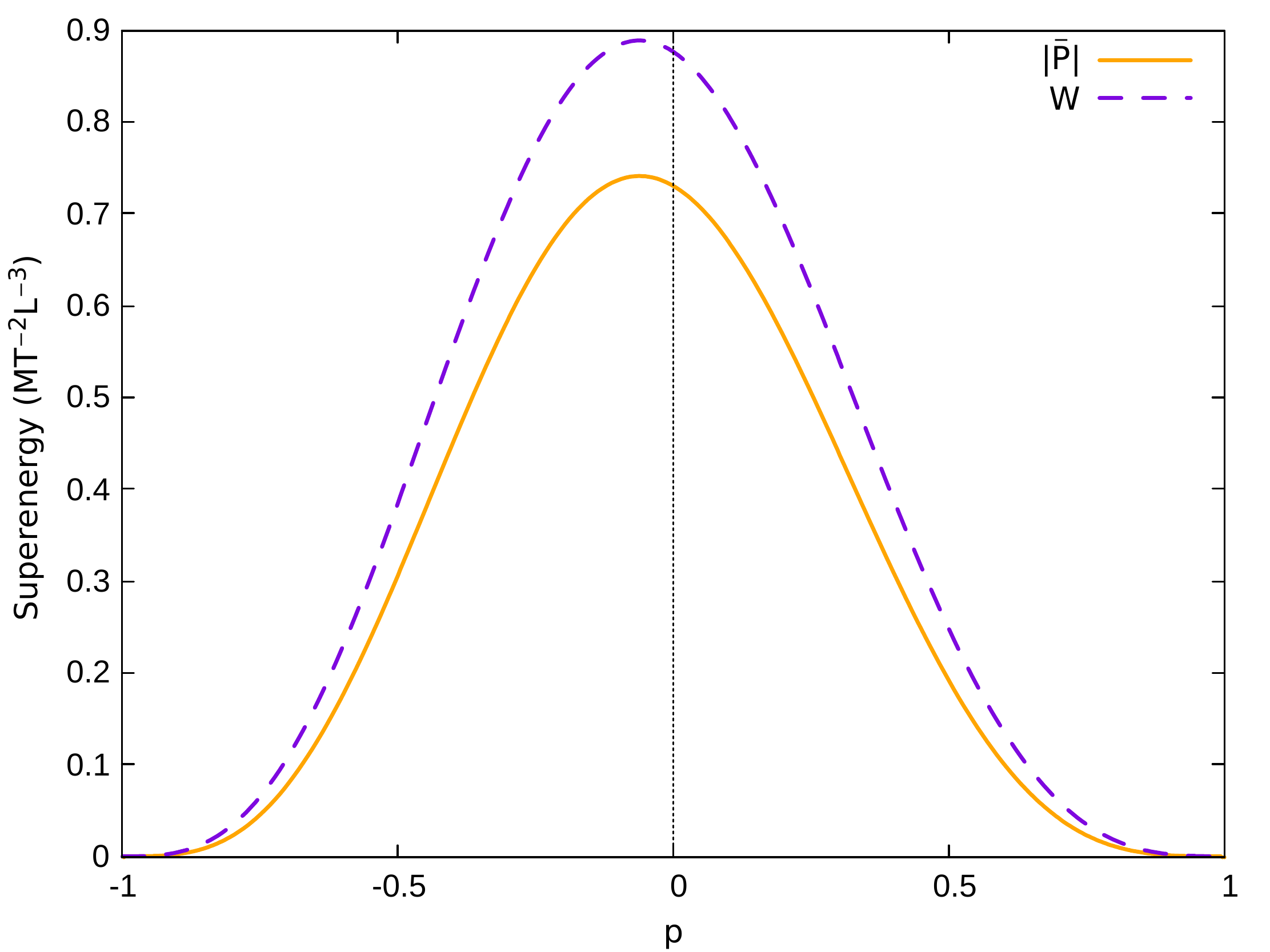}
		 	\caption{Asymptotic canonical superenergy for the C-metric with a choice of parameters $ \Lambda=1 $, $ \alpha=m=1/4 $. The solid line represents the norm of the canonical super-Poynting vector of \cref{super-PinC}; observe that it is non-vanishing in the whole range of the coordinate $ p $ --excluding the conical singularities--, indicating the presence of gravitational radiation there. The dashed line corresponds to the canonical superenergy density of \cref{WinC} that, of course, is non-vanishing.}\label{figure}
		 \end{figure}
		 To measure the effects of $\Lambda$ with respect to the asymptotically-flat case, it is convenient to 
		compute the supermomentum (\ref{super-momentum}) for $\Lambda >0$ and then compare with the radiant supermomentum (\ref{radiant-supermomentum}) for $\Lambda =0$, because of (\ref{eq:limit}). 
		They read, respectively
			\begin{align}
				\ct{p}{^\alpha} &\eqs 2m^2S^{7/2}\brkt{\prn{\alpha^2S+\frac{\Lambda}{3}}\prn{\Lambda+9S\alpha^2}\delta^\alpha_q+\alpha S\prn{2\Lambda+9S\alpha^2}\delta^\alpha_p}\quad, \label{pinC}\\
			\ct{\Q}{^\alpha} &\eqs 18m^2 S^{11/2}\alpha^3\prn{\alpha\delta^\alpha_q+\delta^\alpha_p}\quad.\label{QinC}
			\end{align} 
			Observe that there are linear and quadratic terms in $\Lambda$ in \eqref{pinC} and that it contains a tangent and an orthogonal part to $ \scri $. The former is the asymptotic super-Poynting --the $ \Lambda $-rescaled version of \cref{super-PinC}--, which vanishes if $ \alpha=0 $ and indicates the absence of radiation; the latter, the asymptotic superenergy density that contains not only radiative but also purely Coulomb contributions\footnote{A more elaborate characterisation of the radiative and Coulomb components of the gravitational field will be presented in \cite{Fernandez-Alvarez_Senovilla}.} --accordingly, it does not vanish in general if $ \alpha=0 $. In contrast, the asymptotic radiant supermomentum \eqref{QinC}, which is lightlike and determines the presence of radiation for $ \Lambda=0 $, vanishes if $ \alpha=0 $.

		\end{example}
	\section{Discussion}
 	We have presented a new geometric, covariant and gauge-invariant criterion to characterize gravitational waves at infinity in the presence of a {\em non-negative} cosmological constant.  
	In relevant examples, the criterion gives the expected results. More importantly, we have argued that it is equivalent to the standard one --based on the news tensor-- for the $\Lambda =0$ case. Remarkably, it is entirely based on the information encoded in $(\scri,h_{ab},D_{ab})$ and, using that same data solely, we have investigated a way of dealing with incoming radiation at $ \scri $. This technique induces appropriate further structure on $\scri$ and we are developing a framework \cite{Fernandez-Alvarez_Senovilla} in order to recover other properties of the asymptotically flat case. In particular, ways to define asymptotic symmetries at infinity, as well as  enlightening evidences for the existence of news-like tensor fields on $\scri$ in the presence of a positive cosmological constant, emerge.
			
		\subsection*{Acknowledgments}
		Work supported under Grants No. FIS2017-85076-P (Spanish MINECO/AEI/FEDER, EU) and No. IT956-16 (Basque Government).
	
	\bibliography{radiation_condition_dS.bib}{}
	\end{document}